\documentclass[12pt]{article}

\def\be{\begin{equation}}
\def\ee{\end{equation}}
\def\ba{\begin{eqnarray}}
\def\ea{\end{eqnarray}}
\def\half{{1 \over 2}}
\def\tc{t_{\rm cos}}
\def\mass{{\cal M}}
\def\MPl{M_{\rm Pl}}
\def\Le{|L\rangle}
\def\Ri{|R\rangle}
\def\omin{\omega_{\rm min}}
\def\omax{\omega_{\rm max}}
\def\Dt{\Delta t}
\begin{document}
\title{A dual description of decoherence in de Sitter space}
\author{S. Khlebnikov \vspace{0.1in} \\
{\it \normalsize Department of Physics, Purdue University, West Lafayette, 
IN 47907, USA}}
\date{January 2003}
\maketitle
\begin{abstract}
Decoherence associated with super-Hubble modes in de Sitter space may
have a dual description, in which it is attributed to interaction 
of sub-Hubble modes with an ``environment'' residing just inside the
observer's horizon. We present a version of such description, together 
with some consistency checks, which it is shown to pass. \\
\hspace*{3.9in} hep-th/0301103
\end{abstract}
\section{Introduction}
Properties of de Sitter (dS) spacetime are of considerable interest
due to the important role that near-dS spaces play in inflationary
cosmology and in the theory of accelerating universe. 
Of particular interest is what can be called the quantum 
structure of the dS space, i.e., the structure that is 
supposed to provide us with 
a microscopic explanation of the value of the entropy \cite{GH}.
In the present
paper, we want to explore what can be learned about that structure from 
looking at the behavior of a two-level system in a dS spacetime.
Two-level systems are ubiquitous probes of various types of quantum
environments, due to their susceptibility to {\em decoherence}---the decay, 
as a result of interactions with the environment, of the off-diagonal
components of the system's $2\times 2$ density matrix. 

A simple way to construct a two-level system in de Sitter space is 
to consider
a scalar field with a degenerate
double-well potential, such that the mass scale $\mu$
of the potential is sufficiently large in comparison with the Hubble 
parameter $H$. In this case, 
perturbative amplification of modes does not occur, and the field within 
a given Hubble volume spends most of the time close to one of the minima. 
Any particles created as a result of transitions between the minima are 
rapidly
blown away by the cosmological expansion. 

Transitions between the minima of the potential will occur via tunneling,
but they will be completely incoherent (meaning that no periodic oscillations 
of probability will emerge) \cite{inst}. 
In global coordinates, this decoherence can be
attributed to the super-Hubble modes of the field. On the other hand, an
individual observer in dS spacetime can access only a single Hubble volume,
so it is natural to ask if he can attribute the decoherence to
physics at---or just inside---his own cosmological horizon. The requirement
that such an alternative, or {\em dual}, description exists is analogous
to the ``complementarity principle'' proposed in various forms 
in refs. \cite{BF,DLS}.

In the dual description, a statistical system (the ``environment''), 
which lives on a sphere just inside the observer's horizon, 
will detect transitions 
between the states of the two-level system and cause decoherence in 
the usual way, through system-environment interactions. Only sub-Hubble
modes of the field will be involved. Since the interaction of the field
with the environment will typically be of the form envisioned in the
AdS/CFT correspondence \cite{ads/cft}, it makes sense to explore if, 
in our case, 
a similar correspondence can be used to learn something about the
``environment''. (This proposal, with the boundary located just
inside the horizon, is completely unrelated
to the dS/CFT correspondence discussed in ref. \cite{ds/cft}.)

We will see that in our case an AdS/CFT-like correspondence 
is subject to various
consistency checks, including a rather stringent one, which
follows from the observation \cite{inst} that
decoherence in de Sitter, as measured by the decay of the overlap between 
different evolution histories, is very strong. 
We show that this consistency check is nevertheless passed, 
due to a conspiracy of two factors: the high temperature of the
environment (a consequence of the near-horizon blueshift) and the form
of the environment's spectral density, which turns out to be Ohmic.

Before we proceed to the main point in Sect. \ref{sect:dual}, we briefly
discuss, in Sect. \ref{sect:thin}, tunneling between degenerate minima 
in de Sitter spacetime. Using the thin-wall approximation, we describe
tunneling in the Hamiltonian formalism and compute the amplitude in the
WKB approximation. Comparing the results
with those obtained by the instanton calculus on the Euclidean four-sphere,
we learn about the role of the back-reaction (which the thin-wall
formalism allows us to take into account) and the nature of the turning
points.

Throughout the paper we consider three spatial dimensions (plus time), 
but the construction also works in two.

\section{Tunneling in the two-level system} \label{sect:thin}
Because tunneling between degenerate minima of a heavy field 
in de Sitter space may not be entirely familiar, we begin
with a brief discussion of it here. In this discussion, 
we consider the extreme case, when the mass scale $\mu$ of the
potential is not merely somewhat larger than the Hubble parameter, but
\be
\mu \gg H \; .
\label{limit}
\ee
This allows us to apply the thin-wall approximation. Our subsequent 
results on decoherence do not depend on the limit (\ref{limit}).

Consider a thin-wall bubble with surface tension $\sigma$ nucleating 
in de Sitter space. The problem can be studied using the well-developed
junction formalism \cite{Israel}; we use conventions of ref. \cite{BKT}.
The metric inside the bubble is taken in the form
\be
ds^2 = - (1- H^2 r^2) dt^2 + \frac{dr^2}{1- H^2 r^2} + r^2 d\Omega^2 \; ,
\label{met-in}
\ee
and the metric outside in the form
\be
ds^2 = - f(r') {dt'}^2 + {{dr'}^2 \over f(r')}  + {r'}^2 d\Omega^2 \; ,
\label{met-out}
\ee
with
\be
f(r') = 1- H^2 {r'}^2 - \frac{2k E}{r'} \; .
\ee
Here $\Omega$ is the solid angle, $k$
is Newton's constant, and $E$ is the energy of the bubble.
Because we consider a strictly degenerate potential, the Hubble
parameter $H$ is the same inside and outside the bubble.

Continuity of the metric requires that at the location of the bubble wall
$r=r'$. Away from the wall, the radial coordinates $r$ and $r'$ differ:
in the direction $r$ increases, $r'$ decreases, and vice versa.
The junction conditions lead to an equation of motion for
the wall \cite{Israel}. In our case, that equation can be adapted
from the general form given in ref. \cite{BKT} and reads
\be
E = M (\dot{r}^2 + 1 - H^2 r^2)^{1/2} - k \frac{M^2}{2r} \; ,
\label{eqm}
\ee
where $M= 4\pi \sigma r^2$ is the wall's ``bare'' mass, and dot denotes 
a derivative with respect to the wall's proper time. 

Values of the turning points are obtained from $\dot{r} = 0$. 
Vacuum tunneling, the case of main interest to us, corresponds to $E=0$.
In this case, the turning points are at $r=0$ (no bubble)
and $r=r_c$, where
\be
H^2 r_c^2 = ( 1 + 4\pi^2 k^2 \sigma^2/ H^2 )^{-1} 
\equiv \frac{1}{b^2} \; .
\label{rc}
\ee
The second term in the bracket represents the back-reaction of the wall 
on the metric. If we were to neglect this term, 
we would obtain $r_c =r_H \equiv H^{-1}$, 
that
is the wall of a critical bubble would emerge precisely at the horizon. 
With the back-reaction included, $r_c < r_H$.

We can now compute the amplitude of tunneling in the WKB approximation.
This is most easily done using the canonical formalism, 
where the right-hand side of (\ref{eqm})
is identified with the Hamiltonian ${\cal H}$ of the bubble. As it stands,
${\cal H}$ is written inconveniently in terms of the proper time 
derivative, but it can be rewritten in terms of a canonical 
momentum.\footnote{
The procedure is quite familiar for the case of a dust shell in otherwise
empty space; for a recent discussion, see ref. \cite{DK}. The presence 
of a cosmological constant in our case introduces only minor technical 
complications.}
We define the cosmic time $\tc$ inside the bubble, so that the metric
(\ref{met-in}) becomes
\be
ds^2 = -d\tc^2 + \exp(2H\tc) (d\rho^2 + \rho^2 d\Omega^2 ) 
\label{met-cos}
\ee
where $\rho$ is the comoving radius, and express $\dot{r}$ through the
wall velocity $v = dr/d\tc$:
\be
\dot{r}^2 = \frac{v^2}{1 - (v - Hr)^2} \; .
\ee
Now it can be verified that 
\be
{\cal H} = \sqrt{p^2 + M^2} + Hrp - k \frac{M^2}{2r} \; ,
\label{H}
\ee
where
\be
p = M \frac{v - Hr}{[1-(v-Hr)^2]^{1/2}}
\ee
is the canonical momentum. The latter turns out to be the standard 
expression for a relativistic momentum, except that it appears here in terms
of the relative velocity of the wall and the expanding background, 
$v_{\rm rel} = v - Hr$.

For a given energy eigenvalue ${\cal H} = E$, the momentum as a function of
$r$ reads
\be
p = \frac{-Hr\left( E + k \frac{M^2}{2r} \right) \pm
\left[ \left( E + k \frac{M^2}{2r} \right)^2 - M^2 (1 - H^2 r^2) \right]^{1/2}}
{1- H^2 r^2} \; .
\ee
Classically forbidden regions correspond to negative values of the expression
in the square bracket.
The tunneling amplitude in the WKB approximation is proportional to
$\exp(-S_{\rm WKB})$, where
\be
S_{\rm WKB}(E) = \int_{r_1(E)}^{r_2(E)} dr (1- H^2 r^2)^{-1}
\left[ M^2 (1 - H^2 r^2) - \left( E + k \frac{M^2}{2r} \right)^2 \right]^{1/2}
\; ,
\label{WKB}
\ee
and $r_1$ and $r_2$ are the turning points, for which the expression
in the square brackets vanishes. Recall that $M$ depends on $r$:
$M= 4\pi\sigma r^2$.

For $E=0$, when $r_1 = 0$ and $r_2 = r_c$, the integral in (\ref{WKB})
is readily calculated:
\be
S_{\rm WKB}(0) = \frac{\pi^2 \sigma}{H^3 b} \left( b - \sqrt{b^2-1} \right)^2
\; ,
\label{S_ham}
\ee
where $b$ is defined by (\ref{rc}). Another corollary to eq. (\ref{WKB}) 
is
\be
\left. \frac{\partial S_{\rm WKB}}{\partial E} \right|_{E=0}
= -\frac{1}{H} \arctan \frac{1}{\sqrt{b^2 - 1}} \; .
\label{deriv}
\ee
So, if we were look at tunneling at nonzero temperature, when 
the rate is an integral over $E$ of $\exp[-2S_{\rm WKB}(E) - \beta E]$ 
(with some prefactor), we would find that as long as the temperature 
remains smaller than $1/\beta_{\rm cr}$,
\be
\beta_{\rm cr} = \frac{2}{H} \arctan \frac{1}{\sqrt{b^2 - 1}} \; ,
\ee
vacuum tunneling ($E=0$) is more important than the nearby 
nonvacuum contributions ($E>0$).

We can now compare these results to those from the Euclidean instanton
calculus. A quantitative comparison can only be made for the limiting case
of vanishing back-reaction, since this is the only case for which
the instanton action has been computed in ref. \cite{inst}. In this
limit, $b\to 1$, and the WKB answer (\ref{S_ham}) becomes
\be
S_{\rm WKB}(0) = \frac{\pi^2 \sigma}{H^3} \; .
\label{S1}
\ee

In the Euclidean approach, on the other hand, the instanton is a four-sphere
of radius $r_H$ with the domain wall positioned along an equator. 
This is a complete periodic instanton, in the terminology of ref. \cite{inst}, 
and to obtain the amplitude of tunneling we need the action of a half of it.
The required half is obtained by cutting the four-sphere with a plane
perpendicular to the equator. The boundary produced by this cutting
has the topology of a three-sphere and 
can be viewed as a result of gluing together the spatial metrics of the
patches (\ref{met-in}) and (\ref{met-out}) along the line $r=r'=r_c$.
These two patches therefore correspond to the two turning points of
the periodic instanton (at Euclidean times $\tau = it = 0$ and
$\tau' = it' = \pi$). 

The requisite Euclidean action is now simply the wall tension times half
of the volume of the equator (the limit (\ref{limit}) is implied):
\be
S_{\rm inst} = \frac{\pi^2 \sigma}{H^3} \; .
\label{S2}
\ee
This coincides with the WKB action (\ref{S1}) obtained in the Hamiltonian
approach.

In conformal coordinates $\rho$ and $\eta = -H^{-1} \exp(-H\tc)$, 
with $\rho$ and $\tc$ defined by (\ref{met-cos}), 
the surface of constant $t=t_{\rm nucl}$ corresponds to \cite{inst}
\be
\eta^2 - \rho^2 = H^{-2} \exp(-2H t_{\rm nucl}) 
\equiv \eta_{\rm nucl}^2 \; .
\ee
Thus, $\eta_{\rm nucl}$ can be viewed as the moment of conformal 
time when the {\em center} of the bubble ($\rho=0$) nucleates. 
On the other hand, the bubble wall at $r=r_c$ corresponds to 
$\rho = -H r_c \eta$ and therefore  nucleates at 
$\eta = \eta_{\rm nucl} / \sqrt{1 - H^2 r_c^2}$, i.e., {\em earlier}
than the center (recall that $\eta < 0$).
For weak back-reaction, one can show that by the time the center 
nucleates, the bubble is already of a super-Hubble size.

\section{A dual description of decoherence} \label{sect:dual}
After a bubble nucleates, it begins to expand. The standard way
to follow the subsequent 
evolution of the bubble is to switch from the static
coordinates (\ref{met-in}) to some global, e.g., conformal, coordinates. 
In such global
coordinates, decoherence between the states built near the minima
of the potential---we now refer to these states as $\Le$ and 
$\Ri$---can be attributed to super-Hubble modes of the field
\cite{inst}. On the other hand, in the dual description, decoherence
is attributed to interaction of sub-Hubble modes with an ``environment'' 
located inside the observer's horizon, so
it will be sufficient for our purposes to work
with a single Hubble patch described by the metric (\ref{met-in}).
In this section, it will be convenient for us to analytically continue
to the Euclidean signature, so that $t = -i \tau$ with real $\tau$,
and measure all distances in units of $r_H = H^{-1}$. The metric becomes
\be
ds^2_E = (1-  r^2) d\tau^2 + \frac{dr^2}{1-  r^2} + r^2 d\Omega^2 \; .
\label{met-e}
\ee
Note that $\tau$ is periodic with period $\beta = 2\pi$.

If the environment is placed on a sphere of radius $r_0 < 1$, the metric
on that sphere will be
\be
ds_0^2 = (1- r_0^2) d\tau^2 + r_0^2 d\Omega^2 \; .
\ee
We see that the environment can be considered as being at inverse
temperature
\be
T_0^{-1} = 2\pi\sqrt{1- r_0^2} \; .
\label{T0}
\ee
In what follows we take $r_0$ to be close to
1, so that $T_0$ is very high (we will see that $1/(1-r_0)^{1/2}$
plays the role of an ultraviolet cutoff for the environment).

To cause decoherence between states of a scalar field $\phi$ the environment
has to interact with the field somehow. We assume an interaction
of the form
\be
S_{\rm int} = \int d\tau d\Omega \Phi(\Omega, \tau) 
{\cal O}(\Omega, \tau) \; ,
\label{Sint}
\ee
where
\be
\Phi(\Omega, \tau) = \phi(r_0, \Omega, \tau) 
\ee
is the boundary value of the field, and ${\cal O}$ is some operator
constructed from variables of the environment. In this section, $\phi$
denotes small fluctuations of the scalar field near one of its minima
or, in fact, any other field with similar properties, such as
a component of the metric tensor (a gravitational wave).

Note that the interaction 
(\ref{Sint}) is of the same form as bulk-boundary interactions
envisioned in the AdS/CFT correspondence \cite{ads/cft}. 
In that case, known dynamics of 
fields in the bulk can be used to explore the properties of operators in 
the boundary theory. Here we apply a similar idea to obtain the
two-point correlation function of the operator ${\cal O}$.

The linearized equation of motion for the field $\phi$ (assuming
minimal coupling to gravity) reads
\be
{1\over r^2} \partial_r \left[ r^2 (1-r^2) \partial_r \phi \right]
- {{\hat L}^2 \phi \over r^2} + \frac{\partial_\tau^2 \phi}{1-r^2} 
- \mass^2 \phi = 0 \; .
\label{eqphi}
\ee
Here ${\hat L}$ is the angular momentum, and $\mass$ is the mass of the
field. Similarly to the AdS/CFT case, we require that the corresponding 
bilinear action matches the effective action for $\Phi$ obtained in
the boundary theory:
\be
S[\phi] \sim \int d\tau d\tau' d\Omega d\Omega' \Phi(\Omega, \tau)
\Phi(\Omega', \tau') 
\langle {\cal O}(\Omega, \tau) {\cal O}(\Omega', \tau') \rangle \; .
\label{cor}
\ee
In global coordinates, solutions for a scalar field in the dS background 
are of course well-known \cite{BD}. 
However, these solutions represent traveling 
waves, while to make use of the correspondence (\ref{cor}) we will need
the standing waves. Rather than expand one set of solutions in another,
it will be easier for us to find the requisite properties of the standing
waves directly from eq. (\ref{eqphi}).

Expanding $\phi$ in the eigenfunctions of ${\hat L}^2$ and in
$e^{in\tau}$, with $n=0,\pm 1,\ldots$, and making the change of variables
$x = r/\sqrt{1-r^2}$, we bring (\ref{eqphi}) to the form
\be
\frac{\sqrt{1+x^2}}{x^2} \frac{\partial}{\partial x} 
\left( \frac{x^2}{\sqrt{1+x^2}} \partial_x \phi \right)
- \frac{l(l+1) \phi}{x^2 (1+x^2)} - \frac{n^2 \phi}{1+x^2}
- \frac{\mass^2 \phi}{(1+x^2)^2} = 0 \; ,
\label{eqx}
\ee
$l = 0,1,\ldots$. Note that if $r$ runs from 0 to some $r_0$ close to 1,
then $x$ runs from 0 to large 
\be
R = \frac{r_0}{\sqrt{1 - r_0^2}} \; .
\label{R}
\ee
Thus, the boundary is now at $x=R \gg 1$.

At $x\ll 1$, eq. (\ref{eqx}) reduces to the equation for a field with
mass squared $\mass_n^2 = \mass^2 + n^2$ in a flat tree-dimensional space.
The solution regular at $x=0$ grows at larger $x$.
For large $\mass_n$
this growth is exponential: $\phi = I_{l+1/2}(\mass_n x) /\sqrt{x}$, where
$I_{\nu}$ is the modified Bessel function. So, in the heuristic picture
of the boundary radiating various fields into the bulk, only fields with
sufficiently low masses will be able to reach the observer without 
being exponentially suppressed. At $x \to \infty$, eq. (\ref{eqx}) becomes
\be
{1\over x} \partial_x  \left( x\partial_x \phi \right)
-\frac{n^2}{x^2} \phi = 0 \; .
\label{eqinf}
\ee
Except possibly for some special values of $\mass$ and $l$, the solution
that was regular at $x=0$ will continue to grow at $x\gg 1$.
According to (\ref{eqinf}), the growth is $\phi  \sim x^{|n|}$ 
for $n\neq 0$, and
$\phi \sim \ln x$ for $n=0$. This growth is a reflection
of the coordinate singularity of the original eq. (\ref{eqphi}) near 
the horizon.

Expanding the boundary value $\Phi(\Omega,\tau)$ in $Y_{lm}(\Omega)$ and
$e^{in\tau}$, we can write each component of the field $\phi$ at
$x \gg 1$ as
\be
\phi_{nlm}(x) \approx  \left( \frac{x}{R} \right)^{|n|} \Phi_{nlm}
\label{pow}
\ee
for $n\neq 0$, 
or $\phi_{0lm}(x) \approx \Phi_{0lm} \ln x / \ln R$, for $n=0$.
Substituting these expressions in the bilinear bulk action, we find that 
the main contribution indeed comes from the large $x$ region, with the 
result that
\be
S[\phi] \sim \sum_{nlm} |n| |\Phi_{nlm}|^2 \; .
\ee
In particular, the contribution from $n=0$ is suppressed as $1/\ln R$
and therefore vanishes in the large $R$ limit. 
Note that for this calculation the field was assumed to be canonically
normalized, which can always be achieved by a suitable redefinition.

The correspondence (\ref{cor}) now tells us that
\be
\langle {\cal O}_{nlm} {\cal O}^{\dagger}_{nlm} \rangle \sim |n| \;, 
\label{mom}
\ee
or in the coordinate space, for $|\tau - \tau'| \ll 2\pi$,
\be
\langle {\cal O} (\Omega,\tau) {\cal O} (\Omega',\tau') \rangle
\sim \delta(\Omega - \Omega') \frac{1}{(\tau-\tau')^2} \; .
\label{coord}
\ee
Eq. (\ref{mom}) can be analytically continued to the Lorentzian signature:
\be
G^R_{lm}(\omega) \sim i \omega \; .
\label{lor}
\ee
Note that this analytical continuation gives a {\em retarded} correlator
(as we indicate by superscript $R$). So, the imaginary part of (\ref{lor})
is, up to a numerical factor, the environment's spectral density
in the corresponding channel:
$\rho_{lm}(\omega) \sim \omega$. We see that the spectral density is Ohmic.

Formally, the local in space behavior of (\ref{coord}) is a consequence 
of the angular pieces dropping
out from the asymptotic equation (\ref{eqinf}). So, strictly speaking,
the delta-function in (\ref{coord}) is coarse-grained: if $l$ in 
eq. (\ref{eqx}) becomes of order $R$ (thus involving angular scales of
order $R^{-1}$), eq. (\ref{eqinf}) will no longer apply. Accordingly,
(\ref{coord}) applies only for $|\Omega - \Omega'| \gg R^{-1}$. Still,
since $R$ is large, eq. (\ref{coord}) implies that the correlation length
is short.

Because the result (\ref{lor}) was obtained in the large $R$ limit, we
need to establish how far in the low $\omega$ region it actually holds.
The minimal $\omega$ corresponds to the minimal nonzero energy transfer
that the environment can accept.
On a sphere of radius $r_0 \approx r_H$, the minimal momentum
transfer is of order $r_H^{-1} \sim H$. Applied perpendicular to a
typical momentum of order $T_0 \sim R$, this momentum transfer results 
in the minimal energy transfer of order
\be
\omin \sim H^2 / T_0 \; .
\label{omin}
\ee

\section{Consistency checks}
An immediate consistency check of the above construction is the question
if a correlator like (\ref{lor}), Ohmic in time (frequency) domain and
local in space, can be obtained for a reasonable environment. We notice
at once that this is precisely the behavior characteristic of 
hydrodynamic modes of a thermal system. So, if the scalar couples to
some combination of the energy and pressure 
(and the graviton to the traceless part of the stress tensor),
the form (\ref{lor}) will emerge, with the coefficient in front of $\omega$
proportional to one of the viscosities of the environment.\footnote{
Hydrodynamics of a finite-temperature boundary was studied recently 
\cite{PSS} using a real-time version of the standard 
{\em anti}-de Sitter/CFT correspondence.
It is curious that here we find a connection 
with hydrodynamics for the very 
different (and much less understood) de Sitter case.}

A somewhat unusual property of eq. (\ref{lor}) is its overall magnitude:
as follows from the calculation via (\ref{cor}), the correlator does not 
contain any large factors. 
In particular, powers of $R$ present in (\ref{pow}) 
cancel out those from the integration over $x$. 
This does not impose any bounds on the viscosities, unless one knows
precisely the coefficient of proportionality between the operator
${\cal O}$ and a component of the stress tensor.
However, there is no such freedom in the calculation of decoherence,
and so we arrive at another consistency check.

We now have two calculations of decoherence
for the two-level system considered in the preceding section.
One is the calculation in the global coordinates \cite{inst}, where
decoherence is attributed to super-Hubble modes, and the other is
the dual calculation, where it is attributed to system-environment
interaction. The results should agree. This requirement is nontrivial
because the calculation 
in global coordinates shows that the effect is very large,
while the magnitude of (\ref{lor}), as we have just seen, is quite
modest. 

To quantify what we mean by a large effect, consider the overlap between
the states obtained from one of the basis states of the two-level system,
say $\Le$, via two different evolution histories. 
These are constrained histories, or rather, 
paths in the functional integral:
we prescribe times at which tunneling has occurred but let all the other
degrees of freedom evolve freely. For example, suppose
the first history
is when nothing happens, and the second is when the system tunnels from
$\Le$ to $\Ri$, spends time $\Dt$ there, and then tunnels back. 
Using the corresponding evolution operators,
we can write the overlap as
\be
\langle L| U_2^{\dagger}(t_1, t_2; \Dt) U_1(t_1, t_2) \Le 
\sim \exp[-Q(\Dt)] \; ,
\label{ove}
\ee
where $t_{1,2}$ are some initial and final moments of cosmic time, and
$Q$ will be called the decoherence exponent. The states
$U_1 \Le$ and $U_2 \Le$ are almost indistinguishable within a single
Hubble volume but, once $\Dt$ exceeds some minimal value, 
are very different
at super-Hubble scales: the second state has two expanding bubble
walls, while the first has none. As a result, the overlap (\ref{ove})
(including all modes, both sub- and super-Hubble) 
rapidly decreases to zero as a function of $t_2$.

The observer cannot measure directly the overlap (\ref{ove}) (since
super-Hubble modes are involved), but he can measure 
the probability $P_L(t)$ to remain in the original state $\Le$ after
time $t$.\footnote{
The time $t$ of the static coordinate system (\ref{met-in}) can be written
in terms of $r$ and the cosmic time $\tc$ as $t=\tc -\half\ln (1-r^2)$.
Note that inside a bubble, the metric can be written in the form
(\ref{met-in}) about any point, not necessarily the bubble's center.
So, we assume that the observer is at $r=0$, where $t=\tc$.}
Because of the rapid decay of the overlap (\ref{ove}) 
(and of similar other ones), this
probability is completely incoherent \cite{inst}:
\be
P_L(t) = \half (1+ e^{-\Gamma t}) \; ,
\label{P}
\ee
where $\Gamma$ is the tunneling rate.
Of course, in the dual description, with the environment residing on a
finite-volume sphere (and therefore, in the presence of an ultraviolet
cutoff, having a finite number of degrees of freedom), we cannot expect
a complete decay of the overlap. (Similar points have been made in
ref. \cite{Maldacena} and, in a context quite close to ours, in ref. 
\cite{DLS}.) So, there will be additive corrections to eq. (\ref{P}).
By itself, that is not a problem, since the derivation of (\ref{P})
has used
classical gravity and so cannot exclude a residual overlap
non-perturbative in $1/\MPl$, e.g., of the form
$\exp(-{\rm const.} \MPl^2 / H^2)$ envisioned in refs. 
\cite{Maldacena,DLS}. Still, this is
a large suppression, and it has to be reproduced in the dual approach,
where $Q$ is viewed as a result of interactions with the environment.

High temperature $T_0$ is to rescue: the dual calculation of 
the decoherence exponent involves the
imaginary part of the {\em time-ordered} correlator,
and that is enhanced
relative to (\ref{lor}) by the Bose factor $\coth(\omega/2T_0)$:
\be
Q(\Dt) \sim \sum_{lm} |\Delta \Phi_{lm}|^2 
\int_{\omin}^{\omax} \frac{d\omega}{\omega^2} 
{\rm Im} G^R_{lm}(\omega) (1-\cos \omega \Dt) \coth(\omega/2T_0) \; ,
\ee
where $\Delta \Phi_{lm}$ is the difference in the boundary value of the
field between the states $\Le$ and $\Ri$
(for a review on decoherence due to system-environment interactions,
see ref. \cite{Leggett&al}). For Ohmic
dissipation, one can replace $\coth(\omega/2T_0)$ 
by its high-temperature limit,
so the decoherence exponent is proportional to $T_0 \Dt$ at times
$\Dt \ll 2\pi/\omin$, where $\omin$ is the minimal
frequency (\ref{omin}), and reaches a plateau at larger $\Dt$. From
(\ref{T0}) and (\ref{R}), we see that $T_0 \sim R$, so if we say
that $R \sim \MPl$, then the overlap (\ref{ove})
computed in the dual theory acquires an appropriately strong 
exponential suppression: $Q(\Dt\to\infty) \propto \MPl^2$.

\textheight  8.55in
\section{Conclusion}
Our main result is that an AdS/CFT-like correspondence between de Sitter
bulk and a high-temperature ``environment'' at a near-horizon boundary 
survives a pair 
of most basic consistency checks. First, the correspondence requires
Ohmic form of certain two-point functions in the boundary theory, and
that is consistent with the presence of hydrodynamic modes at the boundary.
Second, the large decoherence in 
a two-level system in de Sitter space 
is reproduced, up to expected corrections, by a dual
calculation in which it is attributed to interaction with the 
environment.

In relating bulk gravitational dynamics to hydrodynamics 
at the boundary, there in no a priori
reason to consider only equilibrium, thermal
boundary environments. One may as well consider nonthermal states, ranging
from nearly thermal (which will hopefully describe near-dS spaces) to highly
nonthermal, such as turbulence. If the correspondence proposed here holds
up under further scrutiny, it would be extremely interesting to obtain
the geometric duals of such nonthermal states.

This work was supported 
in part by the U.S. Department of Energy through Grant DE-FG02-91ER40681 
(Task B).

\end{document}